\documentclass[preprint2]{aastex6}

\slugcomment{Draft Version \today}

\usepackage{mathtools}
\usepackage{epstopdf}
\usepackage{amsmath}
\usepackage{natbib}
\usepackage{multirow}

\newcommand{\msun}{M$_{\odot}$}

\begin{document}

\title{Gas Loss by Ram Pressure Stripping and Internal Feedback From Low Mass Milky Way Satellites}

\author{Andrew Emerick\altaffilmark{1,2}, Mordecai-Mark Mac Low\altaffilmark{2,1,3}, Jana Grcevich\altaffilmark{2}, Andrea Gatto\altaffilmark{4}}

\altaffiltext{1}{Department of Astronomy, Columbia University, New York, NY, USA}
\altaffiltext{2}{Department of Astrophysics, American Museum of Natural History, New York, NY, USA}
\altaffiltext{3}{Institut f{\"u}r Theoretische Astrophysik, Zentrum f{\"u}r Astronomie der Universit{\"a}t Heidelberg, Heidelberg, Germany}
\altaffiltext{4}{Max-Planck-Institute f{\"u}r Astrophysik, Garching, bei M{\"u}nchen, Germany}

\date{\today}

\begin{abstract}
The evolution of dwarf satellites of the Milky Way is affected by the combination of ram pressure and tidal stripping, and internal feedback from massive stars. We investigate gas loss processes in the smallest satellites of the Milky Way using three-dimensional, high resolution, idealized wind tunnel simulations, accounting for gas loss through both ram pressure stripping and expulsion by supernova feedback. Using initial conditions appropriate for a dwarf galaxy like Leo T, we investigate whether or not environmental gas stripping and internal feedback can quench these low mass galaxies on the expected timescales, shorter than 2 Gyr. We find that supernova feedback contributes negligibly to the stripping rate for these low star formation rate galaxies. However, we also find that ram pressure stripping is less efficient than expected in the stripping scenarios we consider. Our work suggests that, although ram pressure stripping can eventually completely strip these galaxies, other physics is likely at play to reconcile our computed stripping times with the rapid quenching timescales deduced from observations of low mass Milky Way dwarf galaxies. We discuss the roles additional physics may play in this scenario, including host-satellite tidal interactions, cored vs. cuspy dark matter profiles, reionization, and satellite pre-processing. We conclude that a proper accounting of these physics together is necessary to understand the quenching of low mass Milky Way satellites.
\end{abstract}

\section{Introduction}

Environmental effects on galaxy evolution have been well studied for massive galaxies in galaxy clusters, both observationally  \citep[e.g.][]{Cayatte1994, Chung2007, Vollmer2010, Abramson2011, Kenney2014}
and with hydrodynamic simulations \citep[e.g.][]{Abadi1999, Quilis2000, RoedigerHensler2005, Tonnesen2007, Kronberger2008, Roediger2009, Ruszkowski2014}. Ram pressure stripping (RPS) of satellite galaxies by halo gas is a well established mechanism for quenching star formation, and is one mechanism responsible for the observed density-morphology relationship in galaxy clusters \citep{GG72, Abadi1999, Poggianti1999}. With the exception of the Large and Small Magellanic Clouds (LMC and SMC), the most massive of the Milky Way's (MW) satellites, all known dwarf galaxies within 300 kpc of the MW remain undetected in \ion{H}{1} \citep{GP09}, and have no observable star formation \citep{McConnachie2012}. This suggests strong, environmental quenching of satellite dwarf galaxies after infall \citep{Wetzel2015}, which has been argued to be produced by rapid RPS after first pericenter passage \citep{Slater2013,Slater2014}. The importance of environmental quenching is further demonstrated by the lack of quenched dwarf galaxies in the field \citep{Geha2012}. Work combining the properties of observed satellites with accretion histories of subhalos in large scale cosmological simulations \citep[e.g.][]{Slater2014, Wheeler2014, Wetzel2015, Fillingham2015} has placed constraints on the required conditions for quenching and associated timescales over the galaxy stellar mass range 10$^4$ M$_{\odot}$ $<$ M$_{*}$ $<$ 10$^{11}$ M$_{\odot}$, complimenting observational work examining quenching in these galaxies \citep[e.g][]{Geha2006, Geha2012, Bradford2015, Weisz2015}. 

Much focus has been placed on the inefficient quenching of more massive dwarf galaxies, 10$^{8}$ M$_{\odot}$ $<$ M$_{*}$ $<$ 10$^{11}$ M$_{\odot}$ \citep[e.g.][]{Wheeler2014}. For the lowest mass dwarf galaxies, M$_{*}$ $<$ 10$^{6}$ M$_{\odot}$, these works suggest that rapid quenching timescales $<$ 2 Gyr are required in order to explain the near unity quenched fraction of galaxies with this stellar mass. Although RPS is a likely candidate for rapid, environmental quenching, this process has yet to be simulated in detail for these low mass dwarfs; simulation work on stripping of dwarfs around the MW has focused on more massive dwarf satellites \citep[e.g.][]{Mayer2006, Nichols2015, Salem2015}. In this work, we use hydrodynamic simulations to examine RPS for these low mass dwarf galaxies to determine its ability to quench these galaxies on short timescales.

In addition, we examine whether or not internal feedback in these dwarf galaxies plays a significant role in the environmental quenching process. Supernova feedback within massive galaxies can drive large scale galactic winds that can aid in the stripping process by driving gas out to large radii where it is more easily stripped \citep{Creasey2013, Girichidis2016}. However, the star formation rate in these tiny galaxies is small, on the order of 10$^{-5}$ M$_{\odot}$ yr$^{-1}$ based on observations of the gaseous, star forming galaxies Leo T, with M$_{*}$ $\sim$ 1.2 $\times$ 10$^{5}$ M$_{\odot}$ \citep{Irwin2007, RyanWeber2008}, and Leo P, with M$_{*}$ $\sim$ 5.7 $\times$ 10$^{5}$ M$_{\odot}$ \citep{Rhode2013}. Although the injection of energy through supernova feedback could, in principle, drive gas out from the shallow potential wells of these galaxies, the low star formation and therefore low supernova rates may prevent substantial outflows \citep{Dercole1999, MacLowFerrara1999, Caproni2015, Melioli2015}. Two-dimensional simulations suggest there may be a galaxy stellar mass regime below which the supernova rate is too low to affect stripping, and above which feedback makes a significant contribution to the quenching process \citep{Gatto2013}. For more massive dwarfs, the necessary rapid quenching of satellite dwarfs may very well only be possible through the combined effects of gas stripping and internal feedback, rather than stripping or feedback alone \citep{Nichols2011}. However, it is unclear whether or not this is the case for the lowest mass dwarf galaxies.

Previous simulation work on the evolution of dwarf satellites around the MW generally focused on more massive galaxies \citep[e.g.][]{Mayer2006, Nichols2015, Salem2015}, were run in two-dimensions at low resolution \citep{Gatto2013}, focused on dwarfs in denser environments - such as groups or clusters - \citep{MoriBurkert2000, Marcolini2003}, or were cosmological \citep[e.g][]{Zolotov2012, Brooks2014, Rafieferantosa2015} and unable to adequately resolve the lowest mass dwarf galaxies. 

We conduct the first set of high resolution, three-dimensional simulations examining the stripping and quenching process in the lowest mass dwarf satellites around the MW. We examine in detail whether or not RPS, in combination with supernova feedback, can strip such galaxies within the $\sim$ 2 Gyr time frame necessary to explain the observed quenched fraction of low mass galaxies. We discuss the implications our results might have on star formation quenching in the lowest mass dwarf galaxies.

In Section~\ref{sec:models} we describe our dwarf galaxy models, simulation physics, and numerical methods. Initial conditions for each of our simulations are discussed in detail in Section~\ref{sec:simulations}. We present our results in Section~\ref{sec:results} and discuss their implications in Section~\ref{sec:discussion}.

\section{Models}
\label{sec:models}
In Section \ref{sec:dwarf galaxy model} we discuss the initial conditions of our dwarf galaxy models that are implemented in both our hydrodynamical simulations and our semi-analytic models. In Section~\ref{sec:numerical methods} we describe our numerical methods, in Section \ref{sec:radiative} we discuss our models of radiative heating and cooling, and in Section \ref{sec:supernova} we discus our prescription for supernova feedback. We discuss the current understanding of the hot halo around the MW in Section \ref{sec:halo}, and in this context, motivate our wind tunnel simulations in Section \ref{sec:wind tunnel}.

\subsection{Dwarf Galaxy Model}
\label{sec:dwarf galaxy model}
Our dwarf galaxy models consist of an initially isothermal, spherically symmetric distribution of gas placed in hydrostatic equilibrium with a static, spherical dark matter potential \citep[][hereafter NFW]{NFW}. The dark matter density distribution is given as 
\begin{equation}
\label{eq:NFW DM profile}
\rho(R) = \frac{\rho_s}{R\left(1+R\right)^{2}},
\end{equation}
where $R = r/r_s$, for the dark matter scale radius $r_{s}$, and $\rho_s$ is the characteristic density given as $\rho_s = \delta_c \rho_{\rm{crit}}$. The parameter $\delta_c$ defines the characteristic density and depends on the concentration parameter $c$, where $c = R_{200}/r_s$ and 
\begin{equation}
\delta_c = \frac{200}{3} \frac{c^3}{log(1 + c) - c/(1 + c)}.
\end{equation}
The critical density of the universe, $\rho_{\rm{crit}}$ is taken at its $z = 0$ value. The virial radius is defined as $R_{200}$, or the point at which the average density of the halo is $200~\rho_{\rm{crit}}$ \citep{BryanNorman1998}. Parameter values defining the dark matter profile are given in Table~\ref{table:dark matter properties}, taking H = 70 km s$^{-1}$ Mpc$^{-1}$. The dark matter potential from the NFW distribution in Equation (\ref{eq:NFW DM profile}) is
\begin{equation}
\label{eq:NFW potential}
\phi_{NFW}(R) = - \phi_{o}^{\rm{NFW}} \frac{\log(1 + R)}{R},
\end{equation}
where the constant $\phi_{o}^{NFW} = 4\pi$G$\rho_s r_s^2$. An isothermal gas in hydrostatic equilibrium with the potential in Equation (\ref{eq:NFW potential}) follows the density distribution
\begin{multline}
\label{eq:NFW gas profile}
\rho(R) = \rho_o \exp\Bigg[-C_{\rm{gas}} \phi^{\rm{NFW}}_{o} \times \\ \left(1 - \frac{\log\left(1 + R\right)}{R}\right)\Bigg] 
\end{multline}
where $C_{\rm{gas}}$ = $\mu m_{\rm{p}}$/($k_{\rm{B}} T_{\rm{dwarf}}$), with $m_{\rm{p}}$ the mass of a proton, $k_{\rm{B}}$ Boltzman's constant, and $\mu$ is the mean molecular mass; $\mu = 1.31$ for the dwarf galaxy gas. In all models, the gas distribution is truncated at some specified radius, $r_{\rm{gas}}$, given in Table~\ref{table:LT}..

The dark matter potential is defined given the dark matter mass, $M_{\rm{DM}}$ ($<$ $r_{\rm{DM}}$), interior to some radius and $r_{\rm{s}}$. We fix the radial extent of the gas profile in each simulation and require pressure equilibrium between the ambient halo medium and the edge of the dwarf galaxy. This is obtained by fixing the hot halo number density and modifying the temperature to establish thermal equilibrium. With these conditions, the gas profile of the dwarf galaxy is described by additionally setting $n_{\rm{o}}$, $r_{\rm{gas}}$, $T_{\rm{gas}}$, and $n_{\rm{halo}}$ (see Table~\ref{table:LT} and Section~\ref{sec:simulations}). Therefore, the halo temperature, $T_{\rm{halo}}$, and the dwarf's total gas mass, $M_{\rm{gas}}$, depend on the values of these parameters. We discuss our parameter choices in more detail in Section~\ref{sec:simulations}.

\begin{table}
 \centering
  \caption{Dark Matter Properties of Dwarf Galaxy Models}
 \begin{tabular}{ c c }
 \hline
 \hline
 Property & Value \\
 \hline
 $M_{\rm{DM}}(r < 300~\rm{pc})$   & $7.3 \times 10^{6}$ M$_{\odot}$ \\
 $M_{\rm{200}}$              & $3.1 \times 10^{8}$ M$_{\odot}$\\
 $R_{\rm{200}}$                     & $13.70$ kpc\\
 $r_{\rm{s}}$                       & $795$  pc\\
 $\rho_{\rm{s}}$            & $1.69 \times 10^{-24}$ g cm$^{-3}$\\
 $\rho_{\rm{crit}}(z = 0)$  & $9.74 \times 10^{-30}$ g cm$^{-3}$\\
 \hline
 \end{tabular}
 \label{table:dark matter properties}
\end{table}

\subsection{Numerical Methods}
\label{sec:numerical methods}
\label{sec:hydro}
We use the Eulerian, grid-based, adaptive mesh refinement code \textsc{FLASH v4.2} \citep{FLASH} for our simulations. \textsc{FLASH} is a multi-dimensional hydrodynamics code that includes multiple solvers. We choose to use the Godunov based, Roe-type, split solver to approximately solve the Riemann equations. The gas in our simulations obeys a polytropic equation of state with fixed adiabatic index $\gamma$ = 5/3. Most previous work on gas loss processes for dwarf galaxy satellites has used either two-dimensional grid codes \citep[e.g.][]{MoriBurkert2000,Gatto2013}, or smooth particle hydrodynamics \citep[e.g.][]{Mayer2006}; few have employed a three-dimensional grid-based code \citep[e.g][]{Salem2015}. Our simulations are all run in three dimensions at high resolution.

To prevent numerical errors from shocks colliding with the simulation boundaries, our dwarf galaxies are placed in a large 20$\times$10$\times$10 kpc rectangular box on a 128$\times$64$\times$64 root grid. This corresponds to a base resolution of 156.25 pc. The dwarf is centered 2 kpc from the left, inflowing boundary in the center of a high resolution region of the domain where we enforce a minimum of 3 levels of refinement to a resolution of 39.06 pc, and allow up to 5 levels of refinement at our fiducial resolution to 9.77 pc. We discuss the effects of resolution on our results with a resolution study in Section~\ref{sec:resolution}.

The high resolution region is centered on the dwarf galaxy with a side length of 1.7 kpc, or 100 pc plus 4 times the initial dwarf gas radius. This limits the amount of stripped tail we fully resolve but saves substantial computing time. We performed tests extending this high resolution box an additional 1.5 kpc downstream of the dwarf galaxy. This did not affect the recovered stripping time in any of the test cases. If it does occur, our initial (1.7 kpc)$^{3}$ high resolution box is sufficient to capture any re-accretion of stripped gas. 

Our analysis of these simulations makes extensive use of the \textsc{yt} analysis toolkit \citep{yt}. We note that our simulation outputs have a time resolution of 10 Myr.

\subsection{Radiative Cooling}
\label{sec:radiative}
Radiative cooling plays an important role in the physics of gas stripping. Including radiative cooling can weaken gas removal due to RPS by more quickly dissipating heating from the initial shock wave. This results in a comparatively cooler and denser ISM that takes longer to strip when compared to runs without radiative cooling \citep{Mayer2006}. However, \cite{TonnesenBryan2009} found that radiative cooling can increase the stripping rate in their disk galaxy simulations, though the total amount of stripped gas remained comparable. In the gaseous wake of the galaxy undergoing stripping, cooling can increase fragmentation, forming dense clumps that can fall back onto the galaxy \citep{Mayer2006}. It is unclear which of these effects will dominate in any given situation, and therefore what the net result of including cooling will be, but it does play an important role in the stripping evolution.

We include the effects of radiative cooling in our simulations following the method described by \cite{Joung2006} and \cite{Joung2009}. The cooling function, $\Lambda$(T), follows the radiative cooling curve from \cite{SutherlandDopita} at $T > 2\times 10^4 ~K$, assuming an optically thin plasma at cosmic abundance ($Z/Z_{\odot} = 1$) and equilibrium ionization. For low temperatures, $T \le 2\times 10^4$~K, cooling is adopted from \cite{DalgarnoMcCray1972}, assuming an ionization fraction of 10$^{-2}$. In the presence of realistic distributed photoelectric heating, this cooling curve would yield a thermally stable regime at $10^4 K \le T < 1.7\times 10^4$~K \citep[see Figure 1 of ][]{Joung2006}. Gas below 10$^4$~K would cool until the next thermally stable regime at around 40~K. However, we instead adopt the common practice of using a cooling floor as a simple means to mimic the effects of a distributed heating rate within the ISM of the dwarf galaxy, preventing runaway overcooling of gas within the galaxy without having to explicitly model heating from various sources such as photoionization and photoelectric heating from stars and heating from a metagalactic UV background. This floor is set to the initial gas temperature of the dwarf galaxies, $T_{\rm{min}} = T_{\rm{gas}}$.

\subsection{Supernova Feedback}
\label{sec:supernova}
Our simulations adopt a supernova implementation used previously to model the turbulent evolution of the ISM \citep{Joung2006, Joung2009, Hill2012}. We summarize the method of energy injection here, as implemented previously, and discuss our implementation of the core collapse and Type Ia supernova rates. Each supernova explosion, regardless of type, injects 10$^{51}$ ergs of thermal energy into the volume surrounding the supernova's location. To prevent overcooling, we choose the mass of the heated gas to be small enough to ensure that the initial temperature lies above 10$^{6}$~K, near the minimum in the cooling curve. The energy is injected into the spherical region surrounding the supernova location that encompasses 60 M$_{\odot}$ of gas. In our simulations, supernova have minimum radii of at least 7 pc, up to about 20-30 pc; supernovae occur in our simulations with typical radii around 10~--~14~pc.Within this volume, the $\sim$ 60 M$_{\odot}$ of gas is distributed evenly across the affected grid cells. The 10$^{51}$ ergs are injected over a uniform density distribution in this region.

Our simulations contain no explicit star formation, and therefore no star particles to determine supernova locations and times. Instead, we implement two separate methods for determining the Type Ia and core collapse supernova rates. Type Ia supernovae can occur during and well after periods of star formation in galaxies. To account for this, we implement a constant Type Ia supernova rate based on a two component delay-time function from \cite{Sullivan2006}. The supernova rate is given as 
\begin{equation}
\label{eq:SN1a rate}
R_{\rm{Ia}} = AM_{*}(t) + B\dot{M}_{*}(t)
\end{equation}
where $M_{*}(t)$ is the total stellar mass of the galaxy at a given time, and $\dot{M}_{*}(t)$ is the star formation rate at a given time. \cite{Sullivan2006} finds $A$ = 5.3 $\pm$ 1.1 $\times$ 10$^{-14}$ ($H_{o}$/70)$^2$ SNe yr$^{-1}$ M$^{-1}_{\odot}$ and $B$ = 3.9 $\pm$ 0.7 $\times$ 10$^{-4}$ ($H_{o}$/70)$^2$ SNe yr$^{-1}$ (M$_{\odot}$ yr$^{-1}$)$^{-1}$. We compute $M_{*}$ for each of our galaxies by scaling to Leo T's $M_{*}/M_{\rm{gas}}$ ratio, 0.43 \citep{RyanWeber2008}; and we use Equation (\ref{eq:SFR}) below to compute $\dot{M}_{*}$. The computed, initial value of $R_{\rm{Ia}}$ is used for the entire simulation, with a supernova occurring every $\tau_{\rm{Ia}}$ = 1/$R_{\rm{Ia}}$. The location of the supernova is chosen randomly with an exponentially decreasing probability using a scale radius equal to the adopted stellar extent (see Section~\ref{sec:simulations} for more details), centered on the initial center of the dwarf galaxy and truncated at the dwarf's initial gas radius. 

Short lived, massive stars are the progenitors of core collapse supernovae, so the core collapse rate should correlate with star formation in the dwarf galaxy and, by extension, the cold gas content of the galaxy. Since we do not have explicit star formation, we estimate the core collapse supernova rate using the amount of cold gas bound to the galaxy at a given time. Given that the probability of a supernova occurring in a given time step is $R_{\rm{II}}\times dt$, we use a random number generator to determine whether or not a supernova should occur. We follow a similar method to \cite{Gatto2013} to determine this rate. The Kennicutt-Schmidt law for galaxies \citep{Kennicutt1998}, extended to low surface densities ($<$ 10 \msun\ pc$^{-2}$) in \cite{Roychowdhury2009}, is given as 
\begin{equation}
\label{eq:SFR}
\Sigma_{\rm{SFR}} = \left(2.13 \pm 0.6 \right) \times 10^{-5} \Sigma_{\rm{gas}}^{2.47},
\end{equation}
where $\Sigma_{\rm{SFR}}$ is given in units of \msun\ yr$^{-1}$ kpc$^{-2}$ and $\Sigma_{\rm{gas}}$ in units of \msun\ pc$^{-2}$. We compute the total cold gas surface density in our galaxy at each time step to obtain the star formation rate of the galaxy. This is translated to a supernova rate by assuming a Salpeter IMF \citep{Salpeter1955} using
\begin{equation}
\label{eq:SNII rate}
R_{\rm{II}} \simeq \frac{6 \times 10^{-3}}{{\rm{M_{\odot}}} {\rm{yr}}^{-1}} SFR~{\rm{yr}}^{-1},
\end{equation}
where $SFR$ is in units of \msun\ yr$^{-1}$. If a core collapse supernova occurs, its position is determined by randomly sampling an exponentially decreasing probability distribution centered on the galaxy center and cut off at some radius $r_{*}$, where $r_{*} < r_{\rm{gas}}$, or 300~pc (Table~\ref{table:LT}). In addition, we require that the supernova occurs within a region of cold gas ($T < 2 \times 10^{4}$~K), redrawing a random position until this condition is met. In general, the Type II supernova rate is a factor of several higher than the initial Type Ia supernova rate in our galaxies. However, while the Type II supernovae dominate initially, Type Ia supernovae become the dominant (though infrequent) source of supernova energy injection during the later evolution of the dwarfs, once a substantial amount of cold gas has been stripped.

\subsection{Hot Halo Density}
\label{sec:halo}
\begin{figure}
\centering
\includegraphics[width=0.95\linewidth]{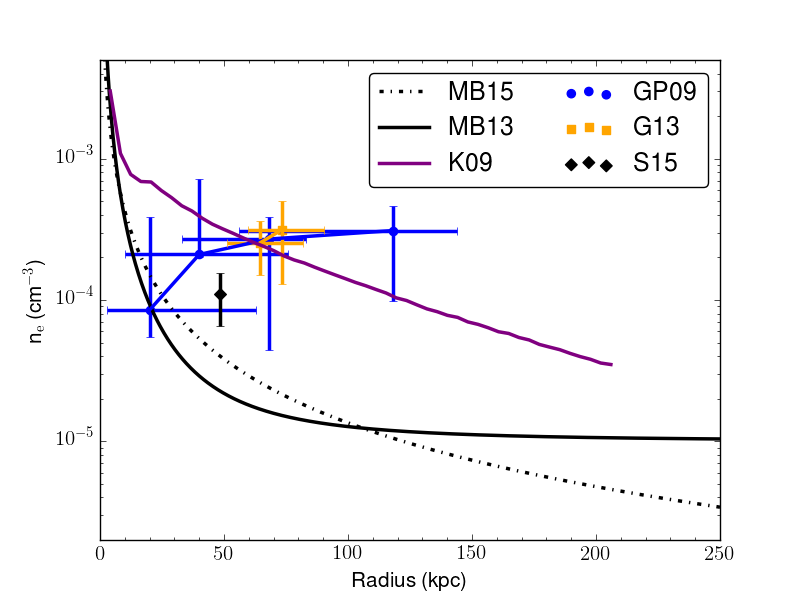}
\caption{Current observational constraints on the number density profile of the MW's hot halo from \cite{MB13} (black, MB13) and \cite{MB15} (black, dash-dotted, MB15). We also include a theoretical model from \cite{Kaufmann2009} (purple, K09), and constraints from both analytic ram pressure stripping arguments with MW dSph's from GP09 (blue, circle), simulations of dSph's from \cite{Gatto2013} (orange, square, G13), and simulations of the LMC from \cite{Salem2015} (black, diamond, S15).}
\label{fig:halo profile}
\end{figure}

Direct observational constraints on the MW's hot halo are limited, probed primarily by \ion{O}{7} and \ion{O}{8} X-ray absorption lines \citep[][MB13 and MB15]{MB13, MB15}. These observations are conducted along a limited number of sightlines, but constrain the radial distribution of the MW's hot gas to a beta model \citep{Makino1998}, which has a power law decay at large radii. Theoretical constraints on the MW's hot halo have been found using RPS arguments for Carina, Ursa Minor, Sculptor, and Fornax dSph's \citep[][GP09]{GP09}, hydrodynamic stripping simulations of Carina and Sextans \citep{Gatto2013}, and stripping simulations of the LMC \citep{Salem2015}. We plot these constraints in Figure~\ref{fig:halo profile} along with a model for the hot halo of the MW developed from hydrodynamic simulations of disk galaxies \citep{Kaufmann2009}. We note the ram pressure arguments provide upper limits and are generally well above the observed profile.

We use a constant wind density and velocity throughout this work. The temperature of the oncoming wind is set by requiring thermal pressure equilibrium between the halo medium and the edge of the dwarf galaxy. This means our wind's thermal properties may be inconsistent with the halo of the MW, but this is sufficient for our purposes as wind temperature does not affect the stripping rate \citep{RoedigerHensler2005, TonnesenBryan2009, Salem2015}. We choose a wind number density of $n_{\rm{halo}} = 10^{-4}$~cm$^{-3}$ for our simulations, in agreement with the constraints provided by \cite{Salem2015} at $r \sim 50$~kpc, but slightly below the upper limit constraints from GP09. 

This density corresponds to pericenter distances of 20~--~30~kpc based on the profiles from MB13 and MB15. Most of the MW hot halo volume has a density lower than this value. In preliminary tests we ran our simulations with wind densities of 10$^{-3}$~cm$^{-3}$ and 10$^{-5}$~cm$^{-3}$ and found that the former produces very rapid stripping in about 0.5~--~1.5~Gyr. The stripping time for the latter, as extrapolated from 2~Gyr of simulation time, is a Hubble time or more. Number densities higher than about $10^{-4}$~cm$^{-3}$ could be reached by orbits that bring the dwarf very close to the MW. We note, however, at this point is is highly likely that tidal stripping may dominate the evolution of the dwarf, possibly completely destroying the satellite.

\subsection{Wind Tunnel}
\label{sec:wind tunnel}
We use wind tunnel simulations in order to mimic the dwarf's motion through its host's hot halo. This allows us to achieve high resolution, allowing us to study these low mass dwarfs without simulating both the entirety of the MW disk or the large volume its orbit would occupy. This has the advantage of both reducing the computational cost of these simulations and allowing us to isolate and examine only the gas stripping physics. Use of wind tunnel simulations to study gas stripping is standard practice for both massive galaxies and dwarf galaxies \citep[e.g.][]{Mayer2006, Gatto2013, Salem2015}. We use a constant wind density and velocity throughout our simulations for better control over the stripping process without having to model variations in cosmologically realistic orbits. With constant density and velocity, our simulations are equivalent to a dwarf satellite on a circular orbit. Alternatively, since a majority of stripping occurs during pericenter of a given orbit, our simulations can be thought of as examining how long a dwarf would have to spend at a given pericenter radius and velocity to be stripped. We test two speeds, $v_{\rm{wind}}$ = 200 km s$^{-1}$ and $v_{\rm{wind}}$ = 400 km s$^{-1}$, directed from the left edge of the simulation box with only a positive $x$ velocity component.

\section{Simulation Initial Conditions}
\label{sec:simulations}

We focus on stripping and feedback in the lowest mass dwarf satellites around the MW, a previously unexplored regime in simulations studying gas stripping. To do this we use observationally motivated initial conditions for our dwarf galaxies. Rather than taking a present day dSph and determining the gas, stellar, and dark matter properties of its pre-infall progenitor, we instead use initial conditions mimicking a current, gaseous low mass satellite of the MW, Leo T. We note that our initial conditions are not meant to be an exact fit to the observed gas and dark matter properties of this galaxy. Leo T is located at $\sim$ 420 kpc and was the first ultrafaint dwarf galaxy detected around the MW \citep{Irwin2007}. It contains both an old stellar population, 6-8 Gyr, and a young population, 200 Myr, suggesting ongoing star formation. It has an observed HI mass interior to 300 pc of $M_{\rm{HI}}$ = 2.8 $\times$ 10$^{5}$ M$_{\odot}$, a stellar mass of $M_{*}$ = 1.2 $\times$ 10$^{5}$ M$_{\odot}$, and a low star formation rate of $1.5-2 \times 10^{-5}$ \msun\ yr$^{-1}$ \citep{RyanWeber2008}. The initial conditions for our two simulated dwarf models are outlined below.

In this work, we fix the dark matter halo for each of our dwarf galaxies, implemented as a static, non-evolving gravitational potential in our simulations; these properties are given in Table~\ref{table:dark matter properties}. Our adopted dark matter mass $M_{\rm{DM}}$($r<300$~pc) $= 7.3 \times 10^{6}$~M$_{\odot}$ is consistent with and bracketed by measurements from \cite{RyanWeber2008}, $M_{\rm{DM}} > 3.3 \times 10^{6}$~M$_{\odot}$, and \cite{Strigari2008}, $M_{\rm{DM}} = 1.30^{+0.88}_{-0.42} \times 10^{7}$~M$_{\odot}$, and consistent with fits from \cite{Faerman2013} with $M_{\rm{DM}} = 6.5 \times 10^{6}$~M$_{\odot}$ for an NFW profile fit, and $M_{\rm{DM}} = 8.0 \times 10^{6}$~M$_{\odot}$ for a Burkert profile fit. We adopt $r_{\rm{s}} = 795$~pc following \cite{Walker2009}, as determined from NFW profile fits to 8 MW dSph's.

%
%

\begin{table*} 
 \centering
 \caption{Dwarf Galaxy Gas and Feedback Properties}
 \begin{tabular}{ c  c  c  c  c  c  c  c }
    \hline
    \hline
    $n_{o}$ (cm$^{-3}$) & $r_{\rm{gas}}$ (pc) & $M_{\rm{gas}}$ (10$^{5}$ \msun) & $T_{\rm{gas}}$ (K) & $r_{*}$ (pc) & $SFR$\textsuperscript{a} (10$^{-6}$ \msun\ yr$^{-1}$) & $\tau_{\rm{SN,1a}}$\textsuperscript{b} (Myr) & $\tau_{\rm{SN,II}}$\textsuperscript{c} (Myr) \\
    \hline
    0.75 & 300.0 & 2.37 & 6.0$\times 10^{3}$ & 170 & 3.90  & 145  & 42.73 \\
    1.50 &       & 4.74 &                    &     & 21.6  & 52.1 & 7.712 \\
    \hline 
 \multicolumn{8}{l}{\textsuperscript{a}\footnotesize{Computed from Equation (\ref{eq:SFR}) given M$_{\rm{gas}}$}} \\
 \multicolumn{8}{l}{\textsuperscript{b}\footnotesize{Fixed rate used in simulation from Equation (\ref{eq:SN1a rate}})} \\
 \multicolumn{8}{l}{\textsuperscript{c}\footnotesize{The estimated initial Type II supernova rate from Equation (\ref{eq:SNII rate}).}}

 \end{tabular}
 \tablecomments{These properties determine the initial conditions for each of our dwarf galaxies. The Type Ia supernova rate is implemented as a constant rate in our simulations, while the Type II rate given is the average expected rate if the dwarf were to evolve in isolation without a wind.}
 \label{table:LT}
\end{table*}

We test two different models for the dwarf galaxy gas profiles that are differentiated by their central gas density, set to either $n_{\rm{o}} = 0.75$~cm$^{-3}$ or $n_{\rm{o}} = 1.50$~cm$^{-3}$. Each dwarf contains an isothermal gas distribution at $T_{\rm{dwarf}} = 6000$~K with an initial gas extent of 300 pc. We note that the resulting peak (central) column density of gas in these two models is $1.76 \times 10^{20}$~cm$^{-2}$and $3.52 \times 10^{20}$~cm$^{-2}$ respectively. Leo T has an observed peak $N_{\rm{HI}}$ of $7 \times 10^{20}$~cm$^{-2}$ \citep{RyanWeber2008}. Thus the gas in our dwarfs is generally more diffuse than in Leo T.

\begin{figure}
\centering
\includegraphics[width=0.95\linewidth]{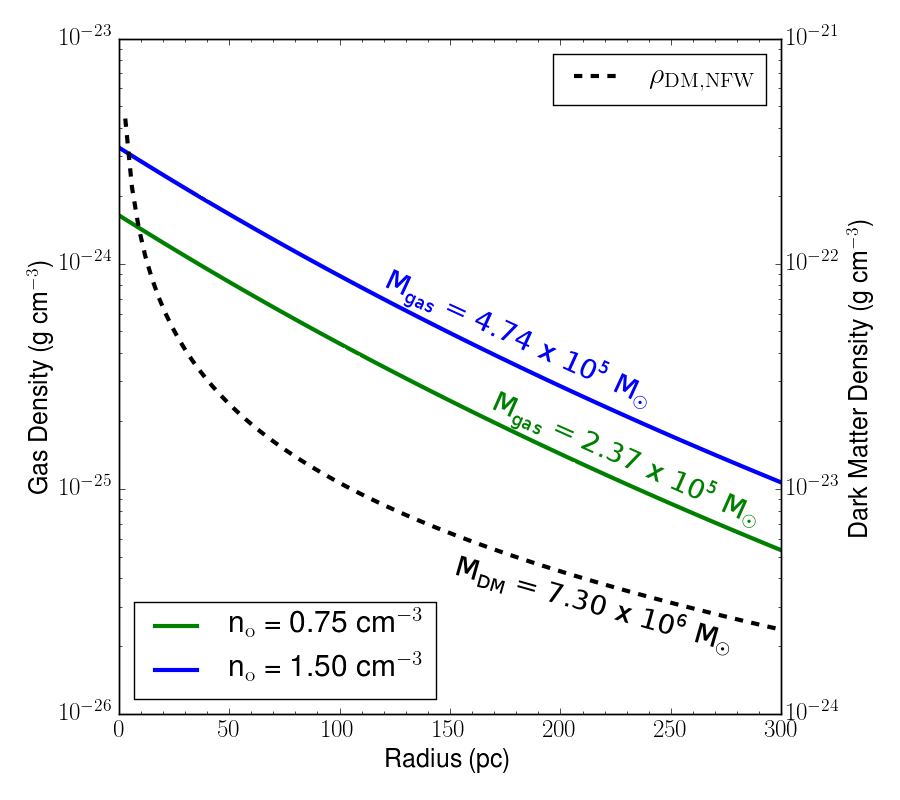}
\caption{Initial gas density (colored, left axis) and dark matter density (dashed, right axis) profiles for the two dwarf galaxy models. The profile and total gas mass interior to $r_{\rm{gas}}$ = 300~pc are shown for the $n_{\rm{o}}$ = 0.75 cm$^{-3}$ (green) and $n_{\rm{o}}$ = 1.50 cm$^{-3}$ (blue) dwarf galaxies, as well as the total dark matter mass interior to 300 pc (black). These properties are summarized in Tables~\ref{table:dark matter properties} and ~\ref{table:LT}.}
\label{fig:LT profiles}
\end{figure}

The total gas mass of the dwarf, $2.34 \times 10^{5}$~M$_{\odot}$ or $4.74 \times 10^{5}$~M$_{\odot}$  (Table~\ref{table:LT}), is determined by these values, by requiring that the dwarf be in hydrostatic equilibrium with its dark matter potential, and by requiring that the edge be in pressure equilibrium with the surrounding gaseous halo. Since we fix the density of the gaseous halo, we find this latter condition by changing the halo temperature; $T_{\rm{halo}} = 2.9 \times 10^{6}$~K and $3.9 \times 10^{6}$~K for the $n_{\rm{o}}$ = 0.75 cm$^{-3}$ and 1.50 cm$^{-3}$ dwarfs respectively. To compute the supernova locations and rates, we must assume some stellar properties of our dwarf galaxy. We again note that we do not include star formation nor the gravitational effect due to a stellar population. We assume instead a stellar radius like Leo T's of $r_{*} = 170$ pc \citep{McConnachie2012}; this is used as the truncation radius in choosing the supernova locations. We use $M_{*} / M_{\rm{gas}} = 0.43$ \citep{RyanWeber2008} for Leo T to set $M_{*}$ in each dwarf to determine the their constant Type Ia supernova rates. These parameters and the resulting supernova rates are given in Table~\ref{table:LT}. Figure~\ref{fig:LT profiles} gives the initial gas and dark matter density profiles for each of our two dwarf galaxies.

\section{Results}
\label{sec:results}
\begin{figure*}
\centering
\includegraphics[width=0.95\linewidth]{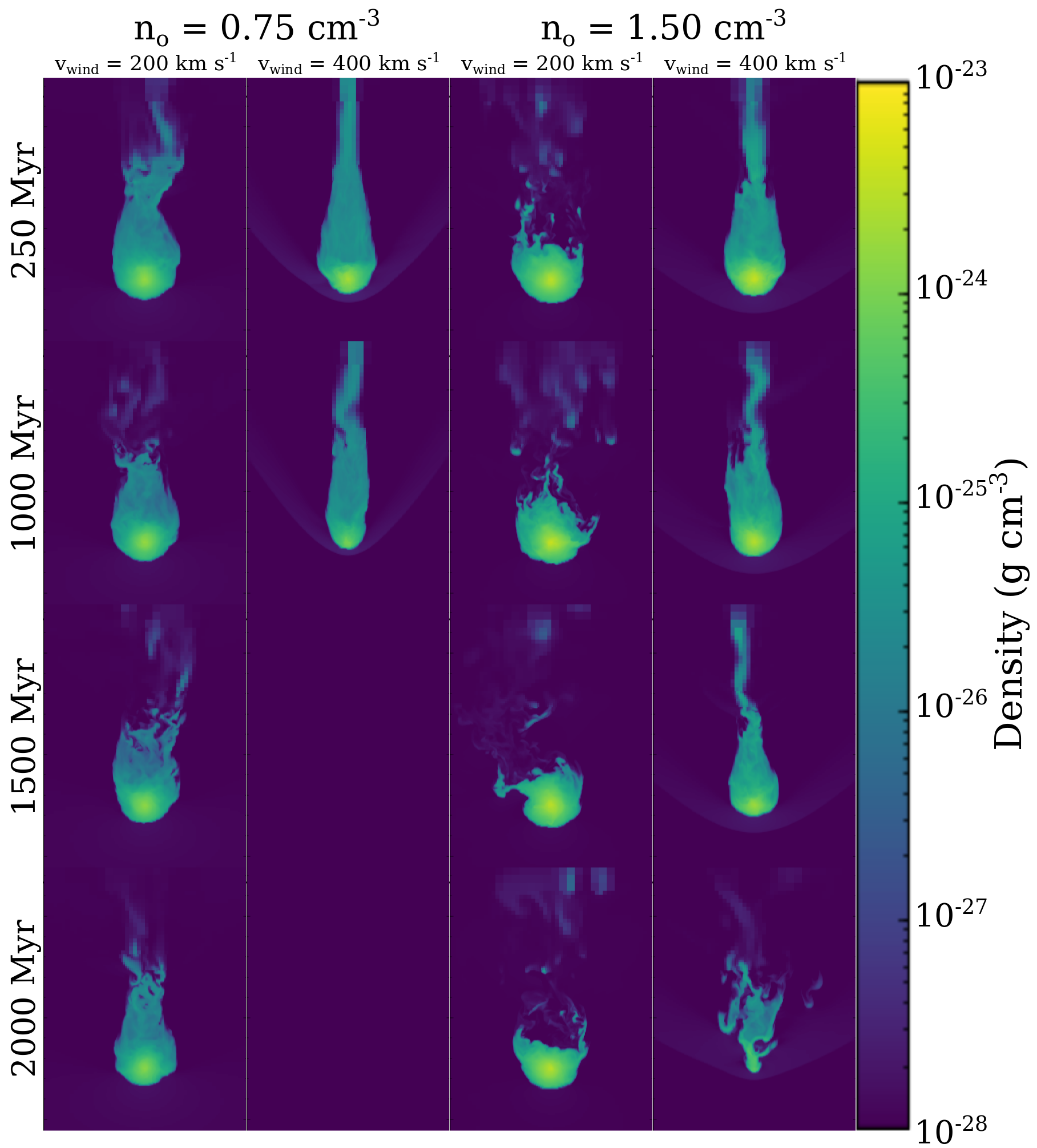}
\caption{Selected density slices at four times for a subset of our simulations, each of the solid black lines in Figure~\ref{fig:simulation results}. The left two columns give the $n_{\rm{o}} = 0.75$~cm$^{-3}$ galaxy run with supernova at the two wind velocities used, while the right columns show the corresponding $n_{\rm{o}} = 1.50$~cm$^{-3}$ simulations. In this orientation, the wind is inflowing from the bottom of each box. We note that these 2~kpc~$\times$~2.6~kpc panels only show a subset of the full simulation volume.}
\label{fig:panel plot}
\end{figure*}

\begin{figure*}
\centering
\includegraphics[width=0.85\linewidth]{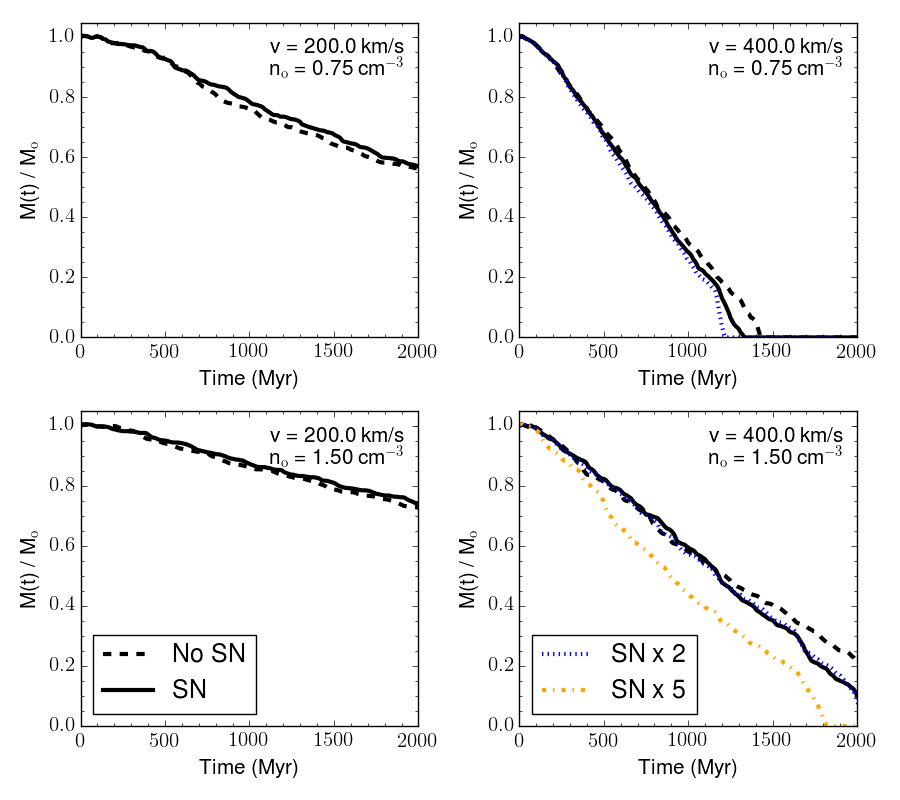}
\caption{The gravitationally bound cold gas mass for each dwarf galaxy run over the 2 Gyr simulation time. For each dwarf model, we run a supernova (black, solid, slices shown in Figure~\ref{fig:panel plot}) and no supernova simulation (black, dashed). We additionally run 3 simulations for the $v_{\rm{wind}}$ = 400~km s$^{-1}$ galaxies where we increase the supernova rate by a factor of 2 (blue, dashed) and 5 (orange, dash-dot). See Table~\ref{table:simulation results} for a list of the obtained and extrapolated stripping times for each simulation.}
\label{fig:simulation results}
\end{figure*}

Figure~\ref{fig:panel plot} shows density slices at four different times during the simulations, comparing the evolution of the two different dwarf galaxies at the two different wind velocities. We show the evolution of the cold gas mass gravitationally bound to the dwarf galaxy over time in Figure~\ref{fig:simulation results}. The solid black lines in Figure~\ref{fig:simulation results} correspond to the density slices in Figure~\ref{fig:panel plot}. The total stripping time and fraction of cold gas remaining at 2~Gyr are presented in Table~\ref{table:simulation results} for each of the lines in Figure~\ref{fig:simulation results}. We use a linear extrapolation from the mass loss over 1.9~--~2.0 Gyr to estimate the stripping time for simulations that still contain gas at 2~Gyr.

Stripping in $\lesssim$~2~Gyr only occurs in the $v_{\rm{wind}} = 400$ km s$^{-1}$ simulations, with the $n_{\rm{o}} = 1.50$ cm$^{-3}$ runs taking just slightly more than 2 Gyr to strip. The extrapolated stripping time for the $v_{\rm{wind}} = 200$ km s$^{-1}$ simulations, however, ranges between about 4~--~7~Gyr; much longer than the 2~Gyr timescale. This significant difference in stripping behaviors between the wind velocities is understandable, as the ram pressure force experienced by these galaxies scales as velocity squared ($P_{\rm{ram}} \propto \rho_{\rm{wind}} v_{\rm{wind}}^2$). In each of these simulations, the stripping rate is fairly low, with a roughly uniform stripping rate over the entire simulation. This generally agrees with the picture of stripping developed in simulations of disk galaxies, where a majority of the evolution time is spent undergoing slow stripping, rather than dominated by rapid, impulsive stripping \citep{Marcolini2003, RoedigerHensler2005, Weinberg2014}. We confirm this in adiabatic simulations of these dwarf galaxies (no radiative cooling and no supernovae) which exhibit a short, initial period of rapid stripping, followed by continuous, low level stripping through the remainder of the simulation.

For these runs, the gas stripping is minimally affected by the inclusion of supernova feedback. This is a different outcome than is seen in massive, disk galaxy simulations, where including supernova feedback plays an important role in driving out gas from the galaxy, assisting stripping. We find here that the supernova rate in these galaxy models is so small, that they only play a minimal role in amplifying the gas stripping rate. We performed three additional tests with supernova (blue and orange/ dotted and dot-dashed lines) where we multiplied the Type II supernova rate by a constant factor (2 or 5), leaving the Type Ia supernovae at the original rate. The SN~$\times$~2 runs still show little difference in evolution until the very end of the simulation. Only the SN~$\times$~5 run shows a significant impact on the gas stripping evolution. Based on visual inspection, the normal supernova rate is infrequent enough that the gas is able to dissipate most of the injected energy from a given supernova before the next one explodes. It is only when the supernova rate increases to the point that multiple supernova contribute to a common superbubble that they are able to drive out gas and assist stripping.

\begin{table}
 \centering
 \caption{Stripping Timescales}
 \begin{tabular}{c  c  l  c  c}
 \hline
 \hline
 n$_{\rm{o}}$ cm$^{-3}$ & v (km s$^{-1}$) & SN Use & f$_{\rm{gas}}$ at 2 Gyr & $\tau_{\rm{strip}}$\textsuperscript{a} (Myr) \\ 
 \hline
 \multirow{5}{*}{0.75} & \multirow{2}{*}{200} & No SN         & 0.56 & (4100) \\\cline{3-5}
                       &                      &    SN         & 0.57 & (4320) \\\cline{2-5}
                       & \multirow{3}{*}{400} & No SN         & - & 1440   \\\cline{3-5}
                       &                      &    SN         & - & 1340   \\\cline{3-5}
                       &                      & SN $\times$ 2 & - & 1220   \\\cline{3-5}
 \hline
 \multirow{6}{*}{1.50} & \multirow{2}{*}{200} & No SN         & 0.73 & (6880)   \\\cline{3-5}
                       &                      &    SN         & 0.74 & (7350)   \\\cline{2-5}
                       & \multirow{4}{*}{400} & No SN         & 0.22 & (2480)   \\\cline{3-5}
                       &                      &    SN         & 0.10 & (2240)   \\\cline{3-5}
                       &                      & SN $\times$ 2 & 0.08 & (2270)   \\\cline{3-5}
                       &                      & SN $\times$ 5 & - & 1830   \\
 \hline
 
 \multicolumn{5}{l}{\textsuperscript{a}\footnotesize{Values in parenthesis are linear extrapolations}}\\

 \end{tabular}
 \tablecomments{Given is the fraction of cold gas bound to the dwarf galaxy at the end of simulation time (2 Gyr) along with the stripping timescale if stripped within this time. If f$_{\rm{gas}} > 0$, we give a parenthesized stripping timescale extrapolated from the last 100 Myr of the simulation.}
 \label{table:simulation results}
\end{table}

\section{Discussion}
\label{sec:discussion}
Here we focus on the role of supernova driven feedback in gas stripping in these dwarf galaxies in Section~\ref{sec:inefficient supernova} and reconcile the stripping timescales seen in our simulations with the expected stripping timescales of observed, low mass dSph's in Section~\ref{sec:reconciling}. Finally, we include a resolution study in Section~\ref{sec:resolution}.

\subsection{Inefficiencies of Supernova Feedback in Low Mass Dwarfs}
\label{sec:inefficient supernova}
Our simulations demonstrate that the supernova feedback in the lowest mass dwarf galaxies is a minor factor in the gas removal process. The star formation rates computed for our dwarf galaxy models are too low to produce enough supernovae to drive substantial outflows. This appears to be because the dwarf's gas has enough time to dissipate the injected supernova energy by the time another occurs. Stacking the effects of multiple supernovae seems to be necessary to drive outflows and affect the gas stripping evolution.

Although \cite{MacLowFerrara1999} showed low mass dwarf galaxies (M$_{\rm{gas}} \sim 10^{6}$ M$_{\odot}$) can be completely destroyed with enough supernova feedback, the supernova rates in our dwarfs are over an order of magnitude below the lowest rate sampled in that work, the lowest of which exhibited a mass ejection fraction of only 20\%. Our results agree with \cite{MacLowFerrara1999}, in that low mass dwarfs cannot drive large outflows from weak supernova feedback. The Carina dwarf galaxy simulations ($M_{\rm{gas}} = 6.3 \times 10^{5}$~M$_{\odot}$) in \cite{Gatto2013} also have too low a supernova rate to affect the gas stripping process. The two lowest mass dwarfs in the \cite{Shen2014} simulation sample have stellar masses of $9.6 \times 10^{4}$~M$_{\odot}$ and $5.3 \times 10^{5}$~M$_{\odot}$, and are unable to drive large outflows. However, it appears that the larger star formation and supernova rates in more massive dwarf galaxies, in spite of a deeper potential well, drive stronger outflows \citep{Shen2014, Caproni2015}. This is seen also in \cite{Gatto2013}, where feedback does play an important role in modifying the stripping rate of their Sextans simulations ($M_{\rm{gas}} = 7.0 \times 10^{6}$~M$_{\odot}$). The lack of observed quenched, isolated dwarf galaxies coupled with their long gas consumption timescales suggests that environmental stripping is necessary for complete gas removal and quenching of dwarf galaxies \citep{Geha2012}. 

Though the exact significance of feedback driven winds in a given galaxy is dependent on the gas geometry, the star formation rate and distribution of star forming regions within the galaxy, and depth of the galaxy's potential well \citep{MacLowFerrara1999}, we further add to the picture of environmental quenching of dwarf galaxies by finding that gas removal in the lowest mass dwarf galaxies must be due to environmental effects alone.

\subsection{Reconciling Inefficient Stripping with the Expected Stripping Timescale}
\label{sec:reconciling}
There has been much recent work examining the quenching processes of dwarf galaxy satellites around the MW and M31 both observationally and theoretically \citep[e.g.][]{Geha2012, Karachentsev2013, Brooks2013, Weisz2015, Wetzel2015, Deason2015, WDGK2015, Fillingham2015, Wheeler2014, Phillips2014, Phillips2015, Slater2013, Slater2014, Slater2015}. Most of this work, however, has focused on dwarf galaxy stellar masses of $10^{8} - 10^{10}$ M$_{\odot}$ where observations indicate that quenching is a highly inefficient process for both satellite and field galaxies. At these masses, roughly 20~--~40\% of observed satellite galaxies are quenched, suggesting quenching timescales around 6~--~9~Gyr. For galaxies of lower stellar mass, about 80\% are quenched at around $M_{*} = 10^{6}$~M$_{\odot}$, and all are quenched for masses down to $M_{*} = 10^{5}$~M$_{\odot}$. Studies examining the observed quenched fraction for these low mass dwarf galaxies (the focus of our work), coupled with accretion histories derived from cosmological dark matter simulations \citep{Slater2014, Wetzel2015, Fillingham2015} suggest efficient quenching timescales, $\leq$ 2 Gyr. Environmental gas stripping is generally assumed as the quenching mechanism for these dwarf galaxies \citep[e.g.][]{Wheeler2014, Slater2014}. However, our work suggests that gas stripping alone, even in combination with internal feedback, is inefficient in the cases we consider and cannot universally quench galaxies within the expected 2 Gyr time frame. 

Our $v_{\rm{wind}}$ = 400 km~s$^{-1}$ simulations have stripping timescales in the range of 1.5~--~2.2~Gyr, comparable to the derived 2 Gyr quenching time. However, this is an optimistic stripping time given that we use a constant 400 km~s$^{-1}$ wind for the entire 2 Gyr simulation. This overestimates stripping in a galaxy with $v_{\rm{peri}} = 400$~km~s$^{-1}$ as a realistic orbit would not result in these wind velocities (or wind densities) for this long. For dwarfs with lower effective average velocities, like our $v_{\rm{wind}}$ = 200 km~s$^{-1}$ simulations, the situation becomes more problematic. These simulations suggest additional physics may be required, beyond what is considered in this work, in order to understand the large quenched fraction of low mass dwarf galaxies. 

Although we find that relying on gas stripping and feedback alone to quench every case of these low mass dwarfs is problematic, these are not the only physical processes that can affect quenching in low mass dwarf satellites of the MW. Tidal interactions between the dwarf and the MW, for example, are often not considered and may be important in the quenching process. We investigate and discuss the role these effects may have in the quenching of these low mass dwarf galaxies below.

\subsubsection{Stripping Under a Weakened Potential}
Recent studies of the kinematics of dwarf galaxies in the Local Group suggest that the dark matter density profiles of dwarfs may contain central cores \citep[e.g.][]{Oh2011} rather than the cuspy profiles expected (see \cite{deBlok2010} for a recent observational review). Assuming galaxies initially contain cuspy profiles, the most likely mechanism of core formation is the gravitational heating of dark matter in the centers of these galaxies through gas motions produced by supernova feedback \citep[e.g.][]{Zolotov2012, Arraki2014, Pontzen2014, Pontzen2015, Nipoti2015, DC14b, Ogiya2015, Read2016}. Core formation through this mechanism is dependent upon the stellar-to-halo mass fraction \citep{DC14b}; too low and feedback is too weak to affect the dark matter distribution, too high and the stars dominate the potential in the centers of the galaxy, preventing the heating of the dark matter. We note here that our Leo T-like models, with M$_{*} = 1.2 \times 10^{5}$ M$_{\odot}$ and M$_{\rm{200}} = 3.1 \times 10^{8}$ M$_{\odot}$, likely have too little supernova feedback to produce significant cores \citep{DC14a,DC14b}.

\begin{figure*}
\centering
\includegraphics[width=0.9\linewidth]{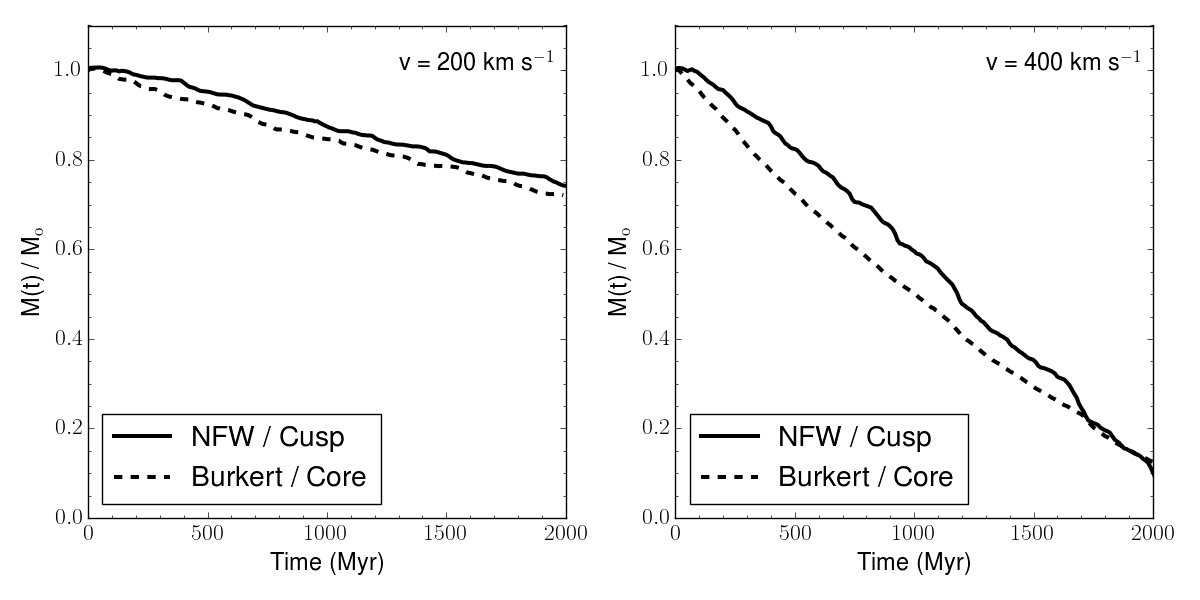}
\caption{Evolution of bound cold gas mass remaining over time. Both models have the same initial gas and dark matter mass within 300 pc, and central gas densities, n$_{\rm{o}}$ = 1.50 cm$^{-3}$. The solid lines give the evolution for a cuspy NFW profile, and dashed for a cored Burkert profile. The latter results in a lower virial mass and overall smaller potential well than the NFW profile, but only reduces the stripping time marginally.}
\label{fig:burkert mass}
\end{figure*}

In our simulations we assume a cuspy, NFW dark matter potential for each dwarf. If cores were to form in these galaxies, it would weaken the gravitational potential and may lead to more rapid stripping. To test this hypothesis, we re-run two $n_{\rm{o}} = 1.50$ cm$^{-3}$ simulations with supernova feedback using a cored, \cite{Burkert1995} profile instead of a NFW profile. The profile was constructed such that it contained the same total gas and dark matter mass interior to 300 pc as our NFW galaxies, with the same central gas density. We present the results of these two simulations in Figure~\ref{fig:burkert mass} (dashed) compared to the equivalent models from Figure~\ref{fig:simulation results} (solid). We find that a cored potential does very little to affect the overall evolution at either wind velocity. The expected stripping times for the v$_{\rm{wind}}$ = 200 km s$^{-1}$ and v$_{\rm{wind}}$ = 400 km s$^{-1}$ simulations are 7.0 Gyr and 2.2 Gyr respectively for the Burkert profiles, and 7.4 Gyr and 2.2 Gyr respectively for the NFW profiles.

\subsubsection{Tidal Effects from the Milky Way}
Our simulations neglect tidal effects the MW might have on gas stripping. They may weaken the stellar and dark matter potentials of our dwarf galaxy, decreasing the stripping timescale. However, the significance of tides strongly depends upon the nature of the dwarf galaxy's orbit and its pericenter distance; being most significant for orbits that plunge deep into the MW potential. Our assumed halo density of 10$^{-4}$~cm$^{-3}$ corresponds to orbital pericenters around 30~kpc, as seen in Figure~\ref{fig:halo profile}. Taking the simple scaling relation for the tidal radius \citep{Johnston1998} $r_{\rm{t}}$ as a function of satellite mass $m$, host mass $M$, and orbital pericenter $r_{\rm{p}}$
\begin{equation}
r_{\rm{t}} = \left(\frac{m}{M}\right)^{1/3} r_{\rm{p}}
\end{equation}
the tidal radius is about 3 kpc for $r_{\rm{p}}$ = 30 kpc and $M_{\rm{MW}} = 10^{12}$ M$_{\odot}$. Since including tidal effects in detail is not possible in our wind tunnel simulations, we approximate tidal interactions by assuming the dark matter distribution of our dwarf galaxy is truncated at $r \geq r_{\rm{t}}$ and unaffected at $r < r_{\rm{t}}$. The reduction in gravitational potential over $r < 2~r_{\rm{gas}}$, or 600~pc, for our dwarfs is roughly uniform at $\sim$20\% of the original. We use this reduction to approximately account for tidal stripping in our simulations by re-running our $n_{\rm{o}} = 1.50$~cm$^{-3}$ supernova simulations with a constant 20$\%$ reduction applied to the dark matter potential. We run an additional simulation using $r_{\rm{p}} = 100$~kpc, corresponding to a 10\% reduction in the gravitational potential over $r < 2~r_{\rm{gas}}$. We note these simulations were not initially in hydrostatic equilibrium, as they were run with the original gas initial conditions in spite of the reduced potential. However, we do not expect this approximation to significantly affect the results, and in any case it will at least give us a lower limit on the stripping timescale.

\begin{figure*}
\centering
\includegraphics[width=0.9\linewidth]{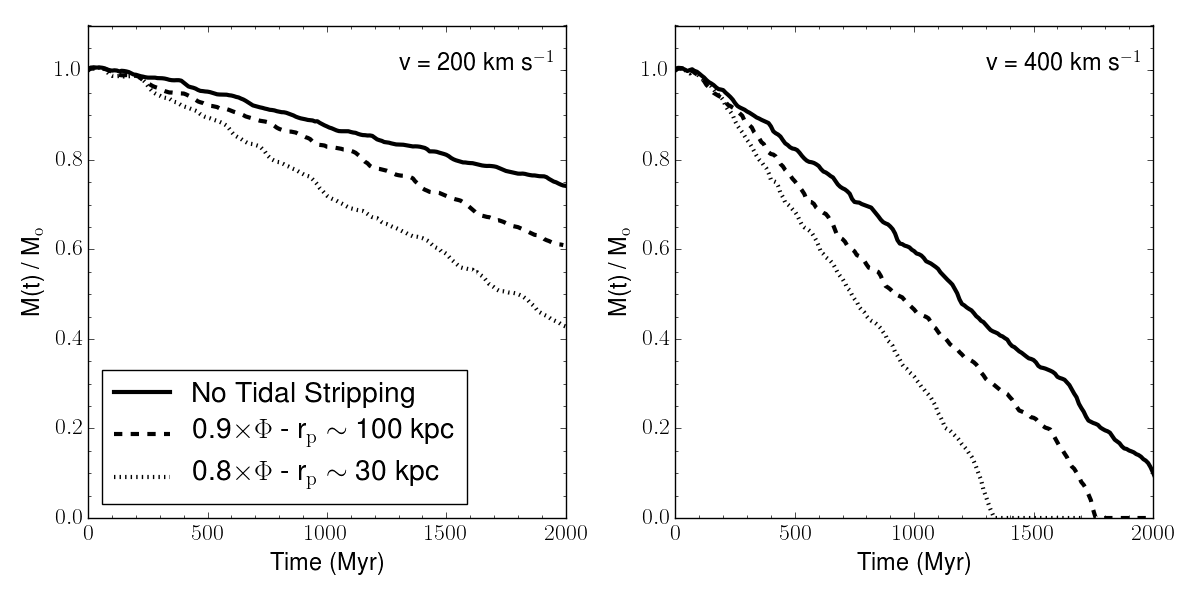}
\caption{Approximate effects of tidal stripping by the MW on our $n_{\rm{o}} = 1.5$ cm$^{-3}$ galaxy. We examined the tidal stripping radius for two pericenter distances, 30 kpc and 100 kpc, dashed and dotted respectively, corresponding to $r_{\rm{tidal}}$ of 5.6 kpc (0.41 R$_{\rm{vir}}$)and 9.4 kpc (0.69 R$_{\rm{vir}}$) respectively. Assuming the galaxy is unaffected at $r < r_{\rm{t}}$, we compute the resulting change in the depth of the potential in each case within $2R_{\rm{gas}}$, or 600 pc. We took the mean change of the potential in each case, which was nearly linear and constant over this range, and re-ran our simulations with a reduced background potential and same initial conditions.}
\label{fig:tidal stripping}
\end{figure*}

The results of this test are shown in Figure~\ref{fig:tidal stripping}, with $r_{\rm{p}} = 100$~kpc (dashed) and $r_{\rm{p}}$~=~30~kpc (dotted) as compared to the original simulations (solid) shown in \ref{fig:simulation results}. As expected, the weakened potentials result in faster stripping in each case. The effect is unable to bring the stripping timescales for the v = 200 km s$^{-1}$ simulation under 2 Gyr, but is significant. The new stripping timescales are 3.5 and 5.0 Gyr for the $r_{\rm{p}}$~=~30~kpc and $r_{\rm{p}}$~=~100~kpc simulations respectively. A comparable fractional decrease is seen in the v = 400 km s$^{-1}$ simulation, yet in this case the tidal effects can reduce stripping to below 2 Gyr, at 1.3 and 1.8 Gyr respectively. Although they may not be relevant in every case, this suggests that accounting for tidal effects between the MW and the infalling dwarf galaxy can be important. Including tidal forces between host and satellite in detail is thus an important area of future study, particularly in capturing cosmologically motivated satellite orbits. Doing so in a cosmological simulation to capture a statistical sample of low mass dwarfs and their orbits is particularly attractive. However, this becomes computationally expensive as it would require the ability to simulate a massive MW-like host galaxy in tandem with its satellites in a hydrodynamic simulation. Although simulations like this have been done focusing on more massive satellites \citep[e.g.][]{Mayer2006}, this is a challenging prospect when the mass of the satellites we are concerned with, $M_{*} \sim 10^{5}$ M$_{\odot}$, are roughly six orders of magnitude smaller than that of the MW.

\subsubsection{Cosmological Orbits and the Halo Density}
\label{sec:cosmological orbits}
We have presented a series of wind tunnel simulations at fixed wind velocity and wind density. In reality, dwarf satellites travel along cosmological orbits encountering a range of wind velocities and densities. Since RPS at pericenter passage dominates, where $P_{\rm RPS} = \rho_{\rm wind} v_{\rm gal}^2$ peaks, stripping in real dwarfs is more impulsive than the continual cases we examine here. It is unclear whether or not this will make stripping more or less efficient in a given case. We consider here the universality of our adopted $P_{\rm RPS}$ compared to expected peak $P_{\rm RPS}$ inferred for satellites of MW mass systems.

\cite{Slater2014} consider accretion histories of subhalos in the Via Lactea 1 and 2 dark matter only simulations to estimate the peak $P_{\rm RPS}$ felt by MW subhalos assuming the \cite{MB13} hot halo density profile shown in Figure~\ref{fig:halo profile}. If RPS is the dominant quenching mechanism, quenching 90\% of MW satellites, they consider the peak $P_{\rm RPS}$ felt by 90\% or more of the Via Lactea halos, calculated as $10^{-14.8}$ dyne cm$^{-2}$. This value can be interpreted to mean that if RPS is the sole cause of gas quenching in MW satellites, it must operate efficiently down to the given $P_{\rm RPS}$. $P_{\rm RPS}$ for our simulations is $10^{-13.4}$ dyne cm$^{-2}$ and $10^{-12.79}$ dyne cm$^{-2}$ for the 200 km s$^{-1}$ and 400 km s$^{-1}$ wind velocities respectively. Our values correspond to the $P_{\rm RPS}$ felt by at least $\sim$75\% and $\sim$50\% of the subhalos in \cite{Slater2014} respectively. Although RPS operates impulsively in satellites with realistic orbits, the time spent at pericenter in a given orbit is on the order of a few hundred Myr, far shorter than the 2 Gyr in our simulations. We expect, then, that our simulations set lower limits for the stripping timescale at the given $P_{\rm RPS}$. Based on the peak $P_{\rm RPS}$ models from \cite{Slater2014}, we argue that at least 25\% of subhalos do not experience sufficient $P_{\rm RPS}$ for efficient quenching, and only at most 50\% experience strong enough peak $P_{\rm RPS}$ to quench within the expected timescales. This generally agrees with a similar analysis presented in \cite{Fillingham2015}.

Finally, uncertainties in the MW's hot halo density are relevant to the universality of our findings. If the MW halo's density is generally higher or has a more shallow fall off with radius than the observations shown in Figure~\ref{fig:halo profile}, our work and previous studies would generally underestimate the typical RPS forces felt by orbiting satellites. Better observational constraints of the MW's halo density profile out to large radii and improved theoretical models of hot halos around MW mass galaxies may help to reconcile the differences between expected stripping timescales and those presented here.

\subsubsection{Group Pre-processing of Low Mass Dwarfs}
Accounting for the complete, cosmological accretion history of a given dwarf satellite may be necessary to recover $<$ 2 Gyr quenching timescales. Tracing the infall histories of subhalos in the ELVIS \citep{ELVIS} suite of cosmological N-body simulations of MW/M31 analogs, \cite{WDGK2015} finds that a substantial fraction of present day subhalos of MW-like galaxies were subhalos of a different pair or group prior to or during infall onto the MW. They find that this ``group preprocessing'' is most prevalent for low mass dwarfs, $M_{*} \leq 10^{6}$ M$_{\odot}$. On average, 50\% of these subhalos were preprocessed. Tidal interactions between dwarfs within these pre-infall groups may play a significant role in the quenching process of these low mass galaxies. They may disturb and weaken the dwarf galaxy potential wells, leading to more efficient stripping upon infall to the MW. Additionally, tidal stirring may heat and puff up the gas within the dwarf, again leading to more efficient stripping. Observational and theoretical work investigating these effects in detail, however, is currently limited. 

Recent work as part of the TiNy Titans study \citep{Stierwalt2015} demonstrates that tidal interactions in dwarf-dwarf pairs drives gas into their outskirts, compared to non-paired dwarf irregulars, as determined by their HI extent \citep{Pearson2016}. If this puffing up of dwarfs is universal in group preprocessing, it would result in more efficient stripping upon infall to a more massive host. Exactly how much this would change the stripping timescales we find is unclear and is an area worthy of future research.

Identifying observational signatures of preprocessing of MW satellites \citep{Deason2015} would be vital to understanding whether or not it played a role for MW dwarfs in particular. Future simulation work could also provide insight into preprocessing. However, doing so in a full cosmological context would again require extremely high resolution hydrodynamic simulations that are able to resolve both these low mass galaxies, $M_{*}$ $\sim$ 10$^{5}$ M$_{\odot}$, and their massive host at the same time. Producing such a simulation is additionally attractive as it would naturally include MW-dwarf tidal interactions post-infall. Although recent work has been able to resolve low mass satellites ($M_{*} > 10^{3}$ M$_{\odot}$) in orbit around a $M_{*} > 10^{6}$ M$_{\odot}$ host \citep{Wheeler2015}, this has yet to be done on the required scales.

\subsubsection{Reionization}
A final possibility is that reionization, rather than RPS or tidal stripping, is the dominant quenching mechanism for low mass dwarf galaxies. This would alleviate any need for the MW environment to remove gas from these galaxies. Inferring infall times from abundance matching in \cite{Rocha2012}, \cite{Weisz2015} compares these to the quenching times derived from the SFH's of MW dwarfs \citep{Weisz2014}. They find that several galaxies with M$_{*}$ $\le$ 10$^{8}$ M$_{\odot}$ were quenched prior to infall onto the MW halo. In fact, a few of these galaxies have been identified as fossils of reionization \citep[See][and references therein]{Weisz2014b}. This points to reionization, rather than environmental stripping, as the quenching mechanism for low mass MW dwarfs. Analysis of \textsc{ELVIS} subhalo accretion histories shows that all subhalos of the MW/M31 analogs were outside the MW/M31 virial radius during reionization, and did not enter it until 2~--~4~Gyr later \citep{WDGK2015}. These authors suggest that the effects of reionization and the MW environment should be separable in observations. Distinguishing the effects of reionization from environmental quenching in these dwarf galaxies through both observation and simulations would be a valuable insight into the evolution of MW satellites. In particular, simulations can answer in detail how reionization affects these galaxies \citep[e.g][]{Simpson2013, CoDa2015}. However, if reionization is the dominant quenching mechanism for these dwarfs, then it must also be able to explain why low mass dwarfs like Leo T and Leo P \citep{Giovanelli2013, McQuinn2015a}, both nearby, low mass dwarfs outside the MW virial radius, contain gas and show observable star formation.

\subsection{Resolution Study}
\label{sec:resolution}
\begin{figure*}
\centering
\includegraphics[width=0.9\linewidth]{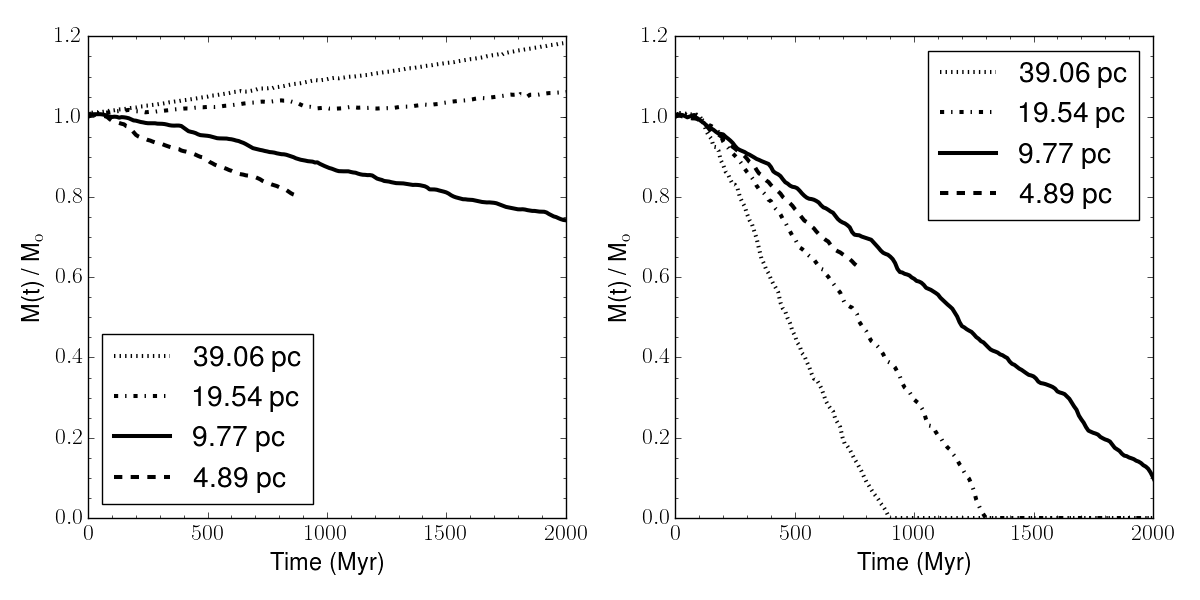}
\caption{Resolution study using the $n_{\rm{o}} = 1.50$~cm$^{-3}$ dwarf galaxy at both $v_{\rm{wind}} = 200$~km~s$^{-1}$ (left) and $v_{\rm{wind}} = 400$~km~s$^{-1}$ (right). Given is our fiducial resolution, 9.77~pc (solid), as compared to two low resolution runs, 19.54~pc (dash-dot) and 39.06~pc (dot), and a partial simulation at higher resolution, 4.89~pc (dashed).}
\label{fig:resolution}
\end{figure*}

We perform the highest resolution three-dimensional simulations of dwarf galaxy RPS to date at our fiducial resolution of 9.77 pc. To understand the effect of resolution in ram pressure stripping simulations of this kind we present a resolution study of the $n_{\rm{o}}~=~1.50$~cm$^{-3}$ dwarf galaxy at both $v_{\rm{wind}}~=~200$~km~s$^{-1}$ and $v_{\rm{wind}}~=~400$~km~s$^{-1}$ in Figure~\ref{fig:resolution}. We present two full low resolution simulations for each, at 39.06~pc and 19.54~pc, and partial runs at 4.89~pc resolution. 

Each velocity case shows a large qualitative change in behavior moving from a resolution of 19.5 to 9.8~pc, corresponding to a change from $\sim 15$ to $\sim 31$ grid cells across the galaxy gas radius. In each case,  there is appreciable gas stripping at a roughly uniform rate for most of the 2 Gyr. Moving to higher resolution, 4.9 pc, shows a quantitative, yet not qualitative change. However, this change is likely too large to consider our 9.8 pc simulations fully converged. Extrapolating the high resolution runs to 2 Gyr, we find that for a factor of two change in resolution, there is only a 41.3\% and 14.6\% change in the recovered stripping times (Table~\ref{table:simulation results}), to 4.31 Gyr and 1.92 Gyr for the 200~km~s$^{-1}$ and 400~km~s$^{-1}$ runs respectively.

The qualitative difference in the behavior moving from low to high resolution for the two different velocities is likely due to the different nature of the gas stripping in each case. While the 200~km~s$^{-1}$ simulation is dominated by low level stripping through fluid instabilities, the factor of four increase in ram pressure force in the $400$~km~s$^{-1}$ run leads to more momentum driven stripping. Unresolved fluid instabilities in the lowest resolution $200$~km~s$^{-1}$ run lead to no stripping, but rather gradual accretion onto the galaxy over time. The low resolution runs in the $400$~km~s$^{-1}$ case contain far too much numerical diffusion, allowing the ram pressure force to more efficiently strip these low resolution galaxies.

This study demonstrates that our stripping times are converging at an acceptable rate in the 200~km~s$^{-1}$ case, suggesting that our standard resolution represents a strong upper limit to the result, but one that appears to be well within a factor of two of the converged result. On the other hand, the 400~km~s$^{-1}$ case may be converged and showing results of random fluctuations \citep{Korycansky2002}, so that one should interpret the result as a final value subject to variation of order 10--20\%. At our fiducial resolution, each 400~km~s$^{-1}$ simulation to 2~Gyr costs on order of 4000~--~5000 CPU hours, or about 3 days on 64 cores.\footnote{We note that the 200~km~s$^{-1}$ simulations are about 25\% cheaper than the 400~km~s$^{-1}$ simulations as the slower wind velocity results in a larger average timestep.} While this is relatively inexpensive, the 4.89~pc simulation with a 400~km~s$^{-1}$ wind cost 30,720 CPU hours (10 days, 128 cores) out to 780 Myr. In total, we estimate the simulations presented in Figures~\ref{fig:simulation results}~--~\ref{fig:tidal stripping} would require $\sim$1.4 million CPU hours at 4.89~pc resolution. Performing multiple simulations at the resolution required for full convergence is computationally expensive. We conclude that our qualitative result, that ram pressure alone is not responsible for low mass satellite quenching, remains valid.

\section{Conclusion}
In this work we present a set of simulations examining the ability of RPS and supernova feedback to remove gas in the lowest mass dwarf galaxy satellites of the MW ($M_{*} \sim 10^{5}$ M$_{\odot}$). In our high resolution wind tunnel simulations, we demonstrate the surprising difficulty of stripping these galaxies, given reasonable assumptions for the orbital velocity and density of the MW's hot halo. Assuming quenching occurs once the galaxy contains no cold gas, this casts doubt on the ability for RPS alone or with supernova feedback to quench these galaxies on the $<$~2~Gyr timescales expected for low mass MW dwarf satellites. We find that the expected supernova rate in these galaxies is too low to effectively influence the stripping rate.  

We conclude that additional physics must be at play, operating in addition to gas stripping, in order to quench these galaxies on relatively short timescales. Tidal stripping effects from interactions between the dwarf and the MW may play a significant role in a fraction of MW dwarfs, but is strongly dependent upon the particular orbits and pericenter distances of the dwarf satellites. In addition, tidal effects during group preprocessing may play a role in either removing gas before infall onto the MW, or puffing up galaxies, allowing for more efficient gas stripping. Finally, re-ionization in the early Universe may have quenched a substantial fraction of these low mass dwarfs, obviating the need for invoking environmental quenching in the first place. However, we do not yet have a complete understanding of which galaxies are quenched during re-ionization, which, if any, galaxies quenched during re-ionization can re-accrete gas and form stars, and what these quenched galaxies would look like at $z = 0$.

Our results demonstrate that RPS does not universally dominate quenching in low mass dwarf satellites. However, this does not mean that RPS is unable to dominate quenching for dwarfs with higher pericenter velocities, which may also experience higher halo densities. As shown in \cite{Slater2014}, approximately half of MW subhalos may have experienced higher RPS forces than we study here. This estimate is dependent upon our understanding of the MW's hot halo density profile, which requires further observations to be well constrained. Modeling each quenching mechanism together with realistic orbits and a range of halo density profiles to estimate the relative importance of each mechanism as a function of satellite stellar mass is a fruitful area of future research.

Disentangling these processes observationally would contribute to a better understanding of the cosmological evolution of the satellites of the MW. In addition, the relative importance of each of these physical processes should be examined further in both semi-analytic models and cosmological hydrodynamics simulations that incorporate tidal effects between host and satellite, the complete accretion history of satellites, group preprocessing, realistic orbits, and the effects of re-ionization in the early Universe. Unfortunately, capturing these physical processes while resolving the lowest mass dwarf satellites around a MW-like host remains computationally challenging, but can be pursued with current semi-analytic models.

\acknowledgements
A.E. is supported by a National Science Foundation Graduate Research Fellowship Grant No. DGE-11-44155. M-MML is partly supported by NSF grant AST11-09395, and the Humboldt Foundation. JG holds a Kathryn W. Davis Postdoctoral Fellowship. AG acknowledges the Deutsche Forschungsgemeinschaft (DFG) for funding through the SPP 1573 ``The Physics of the Interstellar Medium''. The \textsc{FLASH} code was in part developed by the DOE-supported ASC/Alliance center for Astrophysical Thermonuclear Flashes at the University of Chicago. We gratefully recognize computational resources provided by NSF XSEDE through grant number TG-MCA99S024 and Columbia University. Analysis in this work made significant use of the publicly-available \textit{yt} (http://yt-project.org) toolkit. \textit{yt} is the product of a collaborative effort of many independent scientists from numerous institutions around the world. Their commitment to open science has helped make this work possible.

\software{FLASH \citep{FLASH}, yt \citep{yt}}



\bibliographystyle{apj}
\bibliography{msbib}

\end{document}